\documentclass{article}
\usepackage{setspace}

\usepackage{PRIMEarxiv}

\usepackage[utf8]{inputenc} 
\usepackage[T1]{fontenc}    
\usepackage{url}            
\usepackage{booktabs}       
\usepackage{amsfonts}       
\usepackage{nicefrac}       
\usepackage{microtype}      
\usepackage{lipsum}
\usepackage{fancyhdr}       
\usepackage{graphicx}       
\usepackage{bm}
\usepackage{multirow}
\usepackage{adjustbox}
\usepackage{float}
\usepackage{parskip}

\graphicspath{{media/}}     
\RequirePackage{amsthm,amsmath,amsfonts,amssymb}
\RequirePackage[round,authoryear]{natbib}
\RequirePackage[colorlinks,citecolor=blue,urlcolor=blue,linkcolor=blue]{hyperref}

\title{A Bayesian Spatio-Temporal Level Set Dynamic Model and Application to Fire Front Propagation
}

\author{
  Myungsoo Yoo \\
  University of Missouri \\
  \texttt{mym4v@mail.missouri.edu} \\
   \And
  Christopher K. Wikle\thanks{Author of correspondence} \\
  University of Missouri \\
  \texttt{WikleC@Missouri.edu} \\
}

\begin{document}
\maketitle

\begin{abstract}
Intense wildfires impact nature, humans, and society, causing catastrophic damage to property and the ecosystem, as well as the loss of life. Forecasting wildfire front propagation is essential in order to support fire fighting efforts and plan evacuations. The level set method has been widely used to analyze the change in surfaces, shapes, and boundaries. In particular, a signed distance function used in level set methods can readily be interpreted to represent complicated boundaries and their changes in time. While there is substantial literature on the level set method in wildfire applications, these implementations have relied on a heavily-parameterized formula for the rate of spread. These implementations have not typically considered uncertainty quantification or incorporated data-driven learning. Here, we present a Bayesian spatio-temporal dynamic model based on level sets, which can be utilized for forecasting the boundary of interest in the presence of uncertain data and lack of knowledge about the boundary velocity. The methodology relies on both a mechanistically-motivated dynamic model for level sets and a stochastic spatio-temporal dynamic model for the front velocity. We show the effectiveness of our method via simulation and with forecasting the fire front boundary evolution of two classic California megafires – the 2017-2018 Thomas fire and the 2017 Haypress.
\end{abstract}


\section{Introduction}\label{sec:1}
Driven by complex associations from climate warming, changes in land use, and shifts in vegetation type, wildfire plays a vital role in natural ecosystems \citep{brotons}. Although wildfire has some benefits, such as diversifying plants and animals by opening habitat \citep{Pausas}, it often produces significant threats to humans in terms of
property damage and potential loss of life. For instance, the Thomas fire in December 2017, one of the largest wildfires in California history, burned approximately 114,000 hectares of land (440 square miles), damaged over 1,000 buildings, and led to the immediate deaths of one firefighter and one civilian \citep{wan}. To reduce such losses, and to better reduce the cost of mitigating such fires, there is a need for improved models that can more accurately forecast the evolution of wildfires and provide measures of uncertainty on the forecasts.  
That is, accurate forecasting of the boundary of a wildfire may enable immediate fire fighting efforts to focus on the the most critical areas provide more advanced warning for humans to evacuate. It is crucial to develop trustworthy models that can anticipate potential risks from imminent wildfires at the wildland urban interface, and provide uncertainty quantification that enables the creation and evaluation of feasible mitigation and prediction strategies.

Many methods have been developed to model the spread of wildfires, including modeling the Lagrangian movement of tracer particles \citep{clark}, and the method of markers \citep{Filippi}. Lagrangian tracer particles explicitly describe the fire front with four tracer particles within each specified subgrid in the domain of interest. The method of markers represents the boundary by a piecewise linear segment between makers on the line that is the interface between the fire and unburned areas. Those markers are then progressed in the normal direction to the interface line according to a given speed. In some cases, these approaches have been updated in near real time with data to better capture the speed and fire front. For example, \citet{xue} used an ensemble Kalman filter (EnKf) approach to assimilate data into the FARSITE \citep{farsite} fire simulation model. Similarly, \citet{Srivas} used sequential Monte Carlo methods to assimilate data into the DEVS-FIRE fire simulator \citep{devs}. Both \citet{xue} and \citet{Srivas} represented the boundary in terms of points on the boundary. More recently, \cite{green} used deep convolution neural networks with historical fire events to model fire front propagation.

Perhaps the most used approach for modeling the spread of the fire front is the level set approach, where a partial differential equation (PDE) is used to propagate dynamically an implicit level set function. \citet{osher2} first introduced the level set method and \citet{osher} presented a concise introduction and the numerical methods used to solve it. A recent review of these methods and applications can be found in \citet{gibou}. Level set methods have been used in many types of applications. For example, \cite{xie} used level sets to estimate permeability fields, assuming the association between permeability and boundaries. \cite{Iglesias} and \cite{Dunlop} used the level set method in Bayesian inverse problems to better understand subsurface flow. In the context of fire front propagation, \citet{mallet} simplified the wildland fire spread rate proposed by \citet{fendell} and combined it with the level set method to describe the evolution of a fire front. \citet{Rochoux} performed data assimilation with the level set method where the speed of front propagation was controlled by the highly parameterized semi-empirical Rothermel model \citep{rothermel}. \citet{hilton} coupled the level set method with a propagation model of fire spread considering airflow around the fire, which can replicate the behavior of "V" shaped fires observed in experiments. \citet{Alessandri} proposed a method to estimate parameters in the fire spread rate model proposed by \citet{lo}, assuming evolution in direction normal to the fire front. 

Our interest is with using the level set method to predict the evolution of wildland fire fronts in realistic scenarios with spatio-temporally varying speeds learned from the data, and proper uncertainty quantification.  As mentioned above, several studies have combined level set dynamics with a rate of spread parametrization such as the iconic \citet{rothermel} model or those proposed by \citet{fendell} and \citet{lo}. These approaches, which rely on fairly simple semi-empirical parameterizations, are appealing and effective when the environment is homogeneous with known land cover. In other words, when fire is burning over homogeneous fuel sources with relatively steady winds and fairly level terrain, such models function reasonably well. However, they typically are not able to accommodate  intense fires in steep terrain, rapidly changing atmospheric conditions, multiple fuel types, extremely dry conditions, or when the fire itself modifies the local weather.  In addition, the vast majority of fire implementations of level sets treat the problem as deterministic and do not provide a measure of uncertainty in the prediction. However,  \citet{dabrowski} use the level set method combined with a stochastic Rothermel model in a Bayesian representation and ensemble Kalman filter (EnKF) implementation. They are able to update the Rothemel model parameters given observed fire front propagation. They applied their method to a controlled burn with true wind velocity assumed known and constant across the domain. Although this is not realistic for wildland fire propagation, this approach represents the first Bayesian level-set model of fire front propagation.

A comprehensive approach that enables data-driven learning the spatio-temporal varying speed or velocity in the level set method while quantifying uncertainty has not yet been proposed. Therefore, we present a hierarchical Bayesian spatio-temporal level set dynamic model that can accommodate data-driven learning of propagation velocity. The main contribution of our work is as follows. First, we let data determine the propagation speed in the level set method by nesting a stochastic dynamic process within the mechanistically-motivated level-set dynamical model. Thus, we do not need to rely on highly parameterized empirical formulas for the rate of spread in the level set method. Second, the Bayesian framework allows for uncertainty quantification in estimation and forecasting. Given the intense stochastic nature of complex wildfires, providing uncertainty in the forecast will become an essential tool in decision analysis in support of wildfire control. Third, our model can incorporate topographical and landcover covariates into the model for fire front speed. This not only improves forecasting, but provides a formal approach to assess which covariates are important for particular fires. Last, we account for uncertainty in the observed wildfire boundary in our hierarchical representation.  We demonstrate the performance of our model on simulated data and two historical California wildfires, the Thomas fire and the Haypress fire.

This paper is organized as follows. In Section \ref{sec:2}, the data sets that form the motivation and application of this work are described. This is followed by background and a brief description of the level set method in Section \ref{sec:3}. Our Bayesian hierarchical level set dynamic model is described in Section \ref{sec:4}. Sections \ref{sec:5} and \ref{sec:6} are devoted to the application of our method to simulation experiments and real data. Finally, Section \ref{sec:7} provides a brief conclusion and discussion. 

\begin{figure}
\centering
\includegraphics[width=1\linewidth]{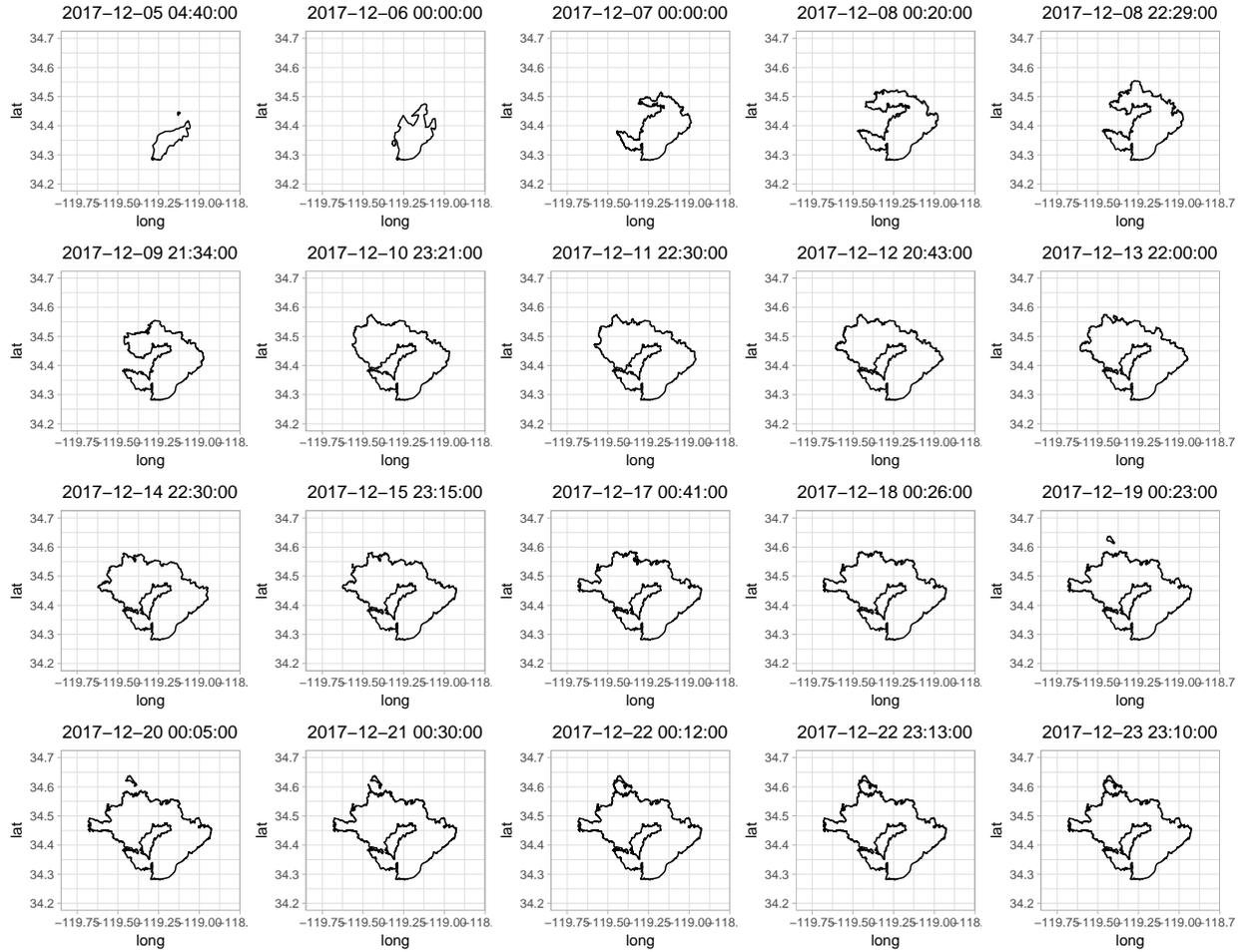}
  \caption{The boundary of the Thomas fire through time. The Thomas fire shows a rapid rate of spread to the north-west at an early stage, then the rate of evolution becomes slower.}
    \label{fig1}
\end{figure}

\section{Data description: The Thomas fire and the Haypress fire}\label{sec:2}
\begin{figure}
\centering
\includegraphics[width=1\linewidth]{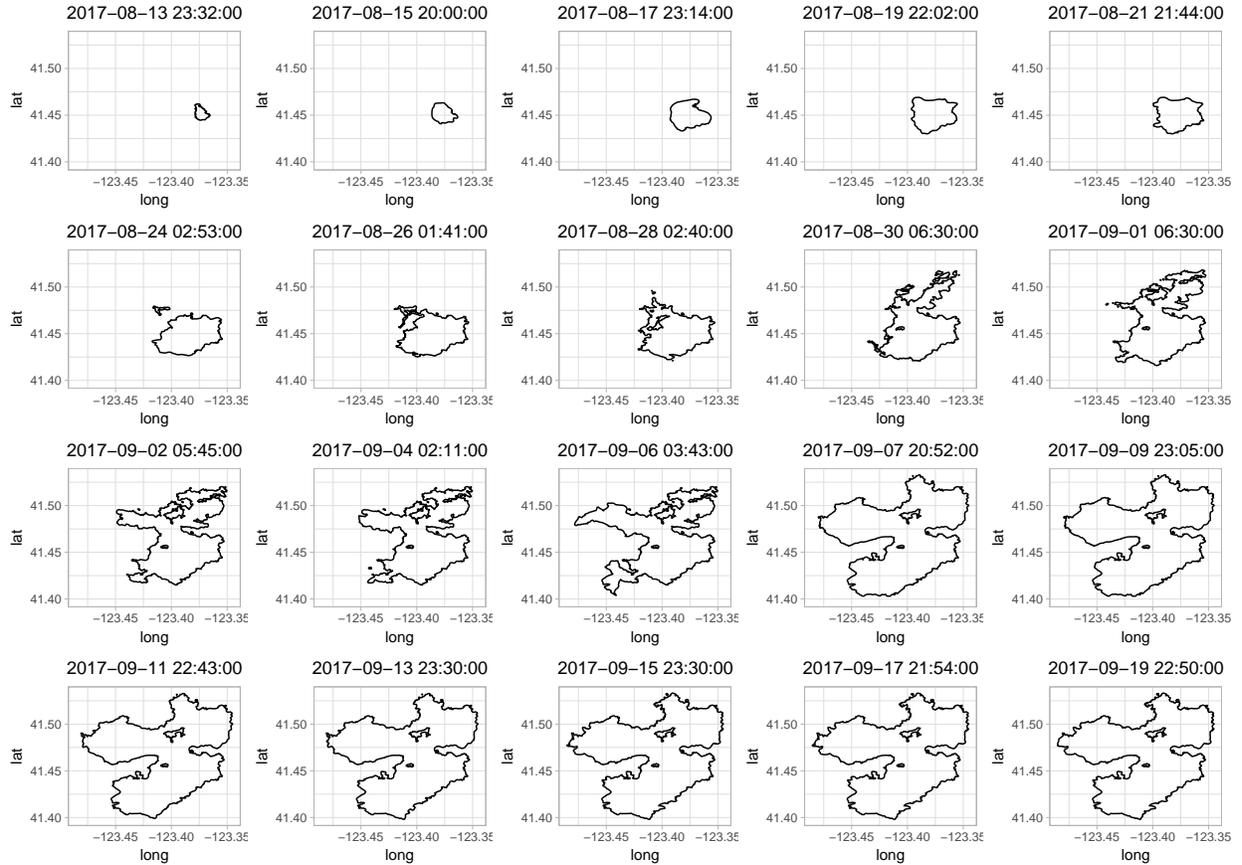}
  \caption{The boundary of the Haypress fire through time. The Haypress fire shows more complicated topological change over time than the Thomas fire.}
    \label{fig2}
\end{figure}
\subsection{The Thomas fire}\label{sec:2.1}
The Thomas fire was a massive wildfire in Southern California between December 2017 and January 2018. After the ignition on December 4, 2017, north of Santa Paula, California, it quickly stretched over the city of Ventura and reached Santa Barbara County. It burned approximately 114,000 hectares of land during its rapid evolution, causing more than a 2.2 billion US dollar loss. Tragically, after containment, 21 fatalities were triggered by post-fire debris flow \citep{addison}. The dataset, obtained as boundaries from the GeoMAC database\footnote{\url{https://wildfire.usgs.gov/geomac/GeoMACTransition.shtml}} \citep{geomac}, consists of twenty-one observations from December 5--24, 2017, from which twenty observations are depicted in Figure \ref{fig1}. The time interval between each measurement ranges from twenty to twenty-five hours. The evolution of the Thomas fire is very rapid to the northwest during the early stages, with the speed decreasing gradually after $t=2017$-$12$-$12$ 20:43:00. 
\subsection{The Haypress fire}\label{sec:2.2}
The Orleans Complex fire in 2017, which occurred in Siskiyou County, California, was ignited in July 2017 and contained in January 2018, having burned over 27,000 acres. Out of nineteen fires comprising the Orleans Complex, the Haypress fire was the largest. We consider twenty-two observations of the fire boundary from the GeoMAC database \citep{geomac}, measured between 2017$-$08$-$13 and 2017$-$09$-$23, with time intervals from forty-four to fifty-three hours (except for one interval of twenty-three hours). As shown in Figure \ref{fig2}, the Haypress fire spreads quickly in the early stage and slows after $t=2017$-$09$-$07$ 20:52:00. The principal distinction between the Thomas fire and the Haypress fire is that the Haypress fire initially showed a rapid stretching to the north and then the primary growth switched towards the west. Another characteristic of the Haypress fire is that new boundaries emerge from spotting and merge as it spreads (e.g., note the separate boundaries at $t=2017$-$08$-$24$ 02:53:00 and $t=2017$-$09$-$01$ 06:30:00 as shown in Figure \ref{fig2}).

\subsection{Spatial varying covariates}\label{sec:2.3}
The "fire behavior triangle," consisting of weather, fuels, and topography, is the main contributor affecting wildfire behavior \citep{Agee}. We use slope and aspect for topography, whereas forest canopy cover and existing vegetation cover are considered proxies for fuels. Forest canopy cover and existing vegetation, obtained from LANDFIRE\footnote{\url{https://landfire.gov/}} \citep{landfire1, landfire2}, describe "the percent cover of the tree canopy in a stand" \citep{landfire1} and "the percent canopy cover by life form" \citep{landfire2}, respectively. Aspect and slope, obtained by R package $\textit{landsat}$ \citep{landsat} and $\textit{elevatr}$ \citep{elevatr}, represent "the steepness of a surface" and "orientation of slope (0: north-facing, 90: east-facing, 180: south-facing, and 270: west-facing)", respectively. The aspect and slope calculations are on the same spatial resolution as the fire boundary locations. However, forest canopy cover and existing vegetation come at finer resolution and are downsampled using bilinear interpolation to match the fire boundary resolution using the function \textit{resample} in R package \textit{raster} \citep{raster}. 
The spatial plots of the covariates are provided in Figures \ref{fig8} and \ref{fig9} in Appendix \ref{appA}. 

\section{Brief background: level set dynamics}\label{sec:3}
This section provides a brief background on signed distance functions and level set dynamics. The level set method, proposed by \citet{osher2}, is based on a mathematical model describing the evolution of an arbitrary dimensional implicit function. We restrict our discussion to $\mathbb{R}^2$ with the implicit function a signed distance function. 
\subsection{Signed distance function}\label{sec:3.1}
One may represent a boundary mathematically through either a Lagrangian or Eulerian formulation. A Lagrangian formulation represents the boundary explicitly through the evolution of specific points on the boundary, whereas an Eulerian formulation defines the boundary implicitly via a contour function \citep{osher}. Although more computationally expensive than Lagrangian methods, Eulerian methods are typically more flexible and have proven effective in representing complex boundaries in wildfires such as those in Figure \ref{fig1} and Figure \ref{fig2}. This advantage becomes more noticeable when the boundary experiences significant topological change through time, such as disjoint boundaries merging into one, or one boundary splitting into multiple boundaries, as is often the case for wildfires. Hence, we consider the implicit approach here.

A signed distance function is often the first choice of implicit functions in level set dynamics as it allows a straightforward representation of complicated boundaries \citep{osher}. A signed distance function $\phi(\bm{s})$ for $\forall \bm{s} \in \mathbb{R}^2$ is defined as
\begin{equation}
\label{eqn:eqn1}
\phi(\bm{s})=
\begin{cases}
-d(\bm{s}), & \bm{s} \in \Omega^{-} \\
0, & \bm{s} \in \partial \Omega \\
d(\bm{s}), & \bm{s} \in \Omega^{+}
\end{cases},
\end{equation}
where $\partial \Omega, \Omega^{+}$ and $\Omega^{-}$ indicate the boundary, exterior, and interior of the boundary, respectively, and the distance function $d(\bm{s})$ for $\forall \bm{s} \in \mathbb{R}^2$ is defined as 
\begin{equation}
\label{eqn:eqn2}
d(\bm{s})=min(\bm{s}-\bm{s_I}),
\end{equation}
for $\bm{s_I}\in \partial \Omega$. To illustrate, consider a unit circle in $\mathbb{R}^2$ centered at $(0,0)$ with a radius of $1$. The corresponding signed distance function is  
\begin{equation}
\label{eqn:eqn3}
\phi(\bm{s})=\sqrt{{s_1}^2+{s_2}^2}-1,
\end{equation}
where $\bm{s}=(s_1,s_2) \in \mathbb{R}^2$. The set of points on the boundary can readily be obtained by $\partial \Omega =\{ \bm{s}:\phi(\bm{s})=0 \}$. Thus, with a signed distance function, one can reduce the task of representing boundaries to finding points $\{\bm{s}:\phi(\bm{s})=0 \}$. Another attribute of the signed distance function is that one only needs to know the sign to determine whether locations are inside or outside the boundary. As we will see, another useful property of the signed distance function is 
\begin{align}
\label{eqn:eqn4}
|\nabla \phi|=1,
\end{align}
where $\nabla \phi$ is spatial gradient of $\bm{\phi}$ and $|\cdot|$ denotes $\ell_2$ norm operation.

\subsection{The level set method}\label{sec:3.2}
The level set method of \citet{osher2} provides a simple, yet effective, approach for modeling the dynamics of an implicit function such as a signed distance function.  A level set equation is defined as the advection-diffusion equation:
\begin{equation}
\label{eqn:eqn5}
\frac{\partial \phi}{\partial t} +\bm{v} \cdot \nabla\phi=0,
\end{equation}
where $\frac{\partial \phi}{\partial t}$ indicates temporal partial derivative of $\phi$, $\nabla \phi$ is spatial gradient of $\phi$, and $\bm{v}=(u,v)^{\top}\in \mathbb{R}^2$ denotes the velocity vector, which depends on time $t$ and spatial location $\bm{s}$. Thus, the direction and speed of evolution depend on the velocity $\bm{v}$ and the spatial gradient. Importantly, (\ref{eqn:eqn5}) can be simplified as follows if the growth is assumed to be in normal direction to the boundary:
\begin{equation}
\label{eqn:eqn6}
\frac{\partial \phi}{\partial t} +v_n \cdot |\nabla\phi|=0,
\end{equation}
where $v_n$ denotes the scalar velocity (speed) in normal direction, with positive and negative speed being outward and inward, respectively. Equation (\ref{eqn:eqn5}) and (\ref{eqn:eqn6}) can be solved numerically, for example, using the so-called ``upwind scheme'' and ``Lax-Friedrichs scheme'', respectively. If $\bm{v}$ or $v_n$ depends on $\phi$, then the level set equation corresponds to a more complicated nonlinear Hamilton-Jacobi equation \citep{osher2}.  

The property of a signed distance function shown in (\ref{eqn:eqn4}) can further simplify the level set equation. Specifically, if $\phi$ is a signed distance function with the flow in the normal direction, (\ref{eqn:eqn6}) can be written very simply as
\begin{equation}
\label{eqn:eqn7}
\frac{\partial \phi}{\partial t} +v_n=0.
\end{equation}
The primary advantage of (\ref{eqn:eqn7}) over (\ref{eqn:eqn6}) lies in the computational efficiency of not needing to evaluate $|\nabla \phi|$. That is, by utilizing a simple forward Euler discretization, (\ref{eqn:eqn7}) can be approximated as 
\begin{equation}
\label{eqn:eqn8}
\frac{\phi_{t+\triangle t} - \phi_t}{\triangle t} +v_n=0,
\end{equation}
where $\triangle t$ is time increment. Therefore, if the normal component of the velocity along the boundary is known, one can easily evolve the boundary to obtain $\phi_{t+\triangle t}$ using this approximation.

Despite the simple formulation in (\ref{eqn:eqn8}), the main challenge in the level set method in applications to wildfire spread is the absence of information about the velocity or the normal direction speed. In addition, the dynamical formulation is quite simple and it is more realistic to account for this simplicity by modeling the level set evolution, conditioned on the velocity/speed, as a stochastic process with additive errors.  We can then focus attention on learning the velocity/speed at a lower level of the model hierarchy. Indeed, a key idea here is to condition the spatio-temporal signed-distance function on a spatio-temporal dynamic process for the normal-direction speed.  This nested combination of a mechanistic dynamic model and the stochastic spatio-temporal process adds significant flexibility and provides a novel data-driven approach to learning fire front propagation.

\section{Bayesian spatio-temporal level set dynamic model}\label{sec:4}
When coupled with a signed distance function, a mechanistic level set equation can naturally be incorporated into a hierarchical Bayesian spatio-temporal dynamic framework  \citep[e.g.,][]{wikle2010}. Bayesian hierarchical models for spatio-temporal dyanmics build dependence conditionally in data, process and parameter stages in a manner that provides formal uncertainty quantification  \citep[e.g., see][for general discussion]{cressie, banejree}. Below we describe these stages for our model.
\subsection{Data model}\label{sec:4.1}
Given the information for the boundary of interest over time, one can obtain an observed signed distance function at $\bm{\tilde{Z}_t}=(\tilde{Z}(\bm{s_1};t),\ldots \tilde{Z}(\bm{s_N};t))^{\top}$  indexed in time $t$ , $1\leq t \leq T$ on $\bm{s}_i\in \mathcal{D} \subset \mathbb{R}^2, 1 \leq i \leq N$ by (\ref{eqn:eqn1}) and (\ref{eqn:eqn2}). A signed distance function with finer resolution (large $N$) can yield a more accurate representation of complicated boundaries, at the expense of computational complexity. Our analysis considers a uniformly distributed grid on $\sqrt{N} \times \sqrt{N}$ a rectangular domain $\mathcal{D}\subset \mathbb{R}^2$ as described for the applications. 

The data model is given by
\begin{equation}
\label{eqn:eqn9}
\bm{\tilde{Z}_t}= \bm{H_t}\bm{\phi_t} + \bm{\epsilon_{t,d}}, \quad \bm{\epsilon_{t,d}} \sim Gau(\bm{0},\sigma^2_d \bm{I}),   
\end{equation}
where $\bm{H_t}$ is a known mapping matrix connecting $\bm{\phi_t}$ to $\bm{\tilde{Z}_t}$, and $\bm{\phi_t}=(\phi(\bm{s_1};t),\ldots,\phi(\bm{s_N};t))^{\top}$ is a hidden spatio-temporal process vector of of interest  at time $t$, which can be interpreted as a "true" signed distance function after filtering out the measurement error. The measurement error vector $\bm{\epsilon_{t,d}}$ is assumed to follow multivariate Gaussian distribution with mean $\bm{0}$ and covariance matrix $\sigma^2_d \bm{I}$. In the case of no missing data, $\bm{H_t}$ becomes the identity matrix $\bm{I}$. It is reasonable with wildfire boundary data to assume that the covariance matrix of $\bm{\epsilon_{t,d}}$ in (\ref{eqn:eqn9}) has a simple (independence) structure as most variation in these types of boundary data can be explained through the process model and the measurement error is thought to be small and uncorrelated in space.

\subsection{Process model}\label{sec:4.2}
The primary process model specifies the mechanistic dynamical evolution of the signed distance function following (\ref{eqn:eqn8}), with the addition of an additive error term, and conditional on the spatial-temporal normal direction speeds at each grid point. That is, vectorizing and rearranging (\ref{eqn:eqn8}), the process model is specified as 
\begin{eqnarray}
\label{eqn:eqn10}
\bm{\phi_{t}}&=\bm{\phi_{t-\triangle t}} - \bm{v_{t-\triangle t}} \triangle t +\bm{\epsilon_{t,p}}, \quad \bm{\epsilon_{t,p}} \sim Gau(\bm{0},\sigma^2_p \bm{I}),
\end{eqnarray}
where $\bm{v_{t-\triangle t}}=(v_{n}(\bm{s_1},t-\triangle t),\ldots, v_{n}(\bm{s_N},t-\triangle t))^{\top}$ is a vector of speed in normal direction to the boundary at time $t-\triangle t$ and $\bm{\epsilon_{t,p}}$ is the process error that is assumed to follow a multivariate Gaussian distribution with mean $\bm{0}$ and covariance $\sigma^2_p\bm{I}$ (note, more complicated spatially-dependent covariance matrices can be specified generally, but our preliminary investigation for the applications considered here suggested the independent assumption was sufficient). It is noted that the property of a signed distance function in (\ref{eqn:eqn4}) may only hold approximately due to discretization, so that $|\nabla \bm{\phi_t}|\approx 1$. The discrepancy induced by this discretization is absorbed in $\bm{\eta_{t,p}}$ and is typically not a concern on the time scales considered here.

In a level set equation, such as (\ref{eqn:eqn6}) and (\ref{eqn:eqn7}), the speed $v_n$ is of primary interest as it governs the evolution of boundaries. Thus, we consider an evolution model for speed that can learn from the data. As we mentioned in Section \ref{sec:1}, this is one of the important contributions of this work. Traditional implementations of level set dynamics in wildfire applications consider an empirical formula for speed \citep[e.g.,][]{rothermel}, that are limited in their ability to learn from the data. It is more appropriate to consider the speed as a stochastic process depending on environmental covariates and a random dynamic process. However, data sets of typical fire progressions do not have a large number of time replicates, limiting the number of parameters that we can estimate for such a dynamic model \citep[e.g., see the discussion in][]{wikle2019}. Thus, we consider a low-rank dynamic  representation, which has been widely used in the context of spatio-temporal statistics \citep{wikle2010_2}, and is typically reasonable since the true dynamics usually exist on a lower dimensional manifold than the observations \citep[see][]{cressie}. Thus, we model the speed $\bm{v_{t-\triangle t}}$ in (\ref{eqn:eqn10}) as a stochastic process given by the mixed effects model
\begin{eqnarray}
\label{eqn:eqn11}
\bm{v_{t-\triangle t}}&=&\bm{X} \bm{\beta}+ \bm{\Psi} \bm{\xi_{t-\triangle t}},
\end{eqnarray}
where $\bm{X}$ is an $N \times P$ matrix of $p$ spatially-varying covariates with coefficients $\bm{\beta}$, and $\bm{\Psi}$ is an $N \times J$ matrix of spatial basis functions with time-varying expansion coefficients $\bm{\xi_{t-\triangle t}}$. Because we require a very low rank process given the relatively few data points in time, we consider bases from a singular value decomposition of a specified covariance function as it provides a low rank basis analogous to empirical orthogonal functions, which are optimal reduction in terms of variance for spatio-temporal data; \cite{cressie}. That is, we first build an exponential spatial correlation function with the range corresponding to about one-third of the spatial domain. Then, $J$ eigenvectors corresponding to the $J$ largest eigenvalues based on the exponential spatial correlation matrix are used as the basis functions. Note, we obtained similar but inferior results with other basis function matrices of the same rank.

Importantly, the low-dimensional expansion coefficients $\bm{\xi_{t-\triangle t}}$ are assumed to follow a first-order vector autoregressive (VAR) process
\begin{eqnarray}
\label{eqn:eqn12}
\bm{\xi_{t-\triangle t}}&=& \bm{M_{\xi}} \cdot \bm{\xi_{t-2\triangle t}}+ \bm{\eta_{t-\triangle t}}, \quad \bm{\eta_{t-\triangle t}} \sim Gau(\bm{0},\bm{\Sigma_{\eta}}),
\end{eqnarray}
where $\bm{M_{\xi}}$ is the transition matrix and $\bm{\Sigma_{\eta}}$ denotes covariance matrix for $\bm{\eta_{t-\triangle t}}$. In our examples, we consider a simpler alternative model with a diagonal transition matrix (because it has fewer parameters to estimate)
\begin{eqnarray}
\label{eqn:eqn13}
\bm{\xi_{t-\triangle t}}&=& diag(\bm{\gamma}) \cdot \bm{\xi_{t-2\triangle t}}+ \bm{\eta_{t-\triangle t}}, \quad \bm{\eta_{t-\triangle t}} \sim Gau(\bm{0},\bm{\Sigma_{\eta}}),
\end{eqnarray}
where $\bm{\gamma}=(\gamma_1,\ldots \gamma_J)^{\top}$ and $diag( \bm{\gamma})$ is a diagonal matrix with $\bm{\gamma}$ the diagonal elements. 

\subsection{Parameter models}\label{sec:4.3}
The last stage of the Bayesian hierarchical model consists of prior distributions for parameters. To facilitate computation, we select the following conjugate prior distributions: 
\begin{eqnarray}
\label{eqn:eqn14}
\sigma^2_{d} &\sim& \text{inverse gamma}(\alpha_d,\beta_d), \nonumber \\
\sigma^2_{p} &\sim& \text{inverse gamma}(\alpha_p,\beta_p), \nonumber\\
\bm{\Sigma_{\eta}^{-1}} &\sim& \text{Wishart}((1000 \bm{I} \times (J- 1) )^{-1},J-1), \nonumber \\
\bm{\beta} &\sim& \text{Gau}(\bm{0},c_{\beta}\bm{I}),\\
\bm{\xi_{0}} &\sim& \text{Gau}(\bm{0},c_{\xi}\bm{I}), \nonumber \\
vec(\bm{M_{\xi}}):=\bm{m_{\xi}} &\sim& \text{Gau}(\bm{0},c_m \bm{I}), \nonumber\\
\bm{\gamma} &\sim& Gau(\bm{0},c_{\gamma} \bm{I}), \nonumber 
\end{eqnarray}
where the $vec(\cdot)$ operator stacks matrix elements columnwise. The initial condition distributions for the VAR models must also be assigned. Here, we assume $\bm{\phi_0}=\bm{\tilde{Z}_1}$ and we denote the initial condition distribution for $\bm{\xi_t}$ as $\bm{\xi_{0}}$. The choice of hyperparameters are application specific and is discussed in Section \ref{sec:5}. 

\subsection{MCMC sampling algorithm}\label{sec:4.4}
Letting $\bm{\Theta}$ denote the set of all parameters and process components,  $\bm{\phi_1},...,\bm{\phi_T}$, the posterior distribution is 
\begin{equation}
\label{eqn:eqn15}
\pi(\bm{\Theta}|\bm{\tilde{Z}_{1:T}}) \propto L(\bm{\tilde{Z}_1,\ldots,\tilde{Z}_T}|\bm{\Theta}) \cdot \pi (\bm{\Theta}),
\end{equation}
where $\bm{\tilde{Z}_{1:T}}=(\bm{\tilde{Z}_1},\ldots,\bm{\tilde{Z}_T})^{\top}$. To draw samples from the posterior distribution, a Gibbs sampling algorithm is used \citep{Ravenzwaaij} as given in Appendix \ref{appB}. The predictive posterior distribution is 
\begin{equation}
\label{eqn:eqn16}
\pi(\bm{\tilde{Z}_{pred}}|\bm{\tilde{Z}_{1:T}})=\int \int \pi(\bm{\tilde{Z}_{pred}}| \bm{\phi_{pred}}, \bm{\Theta} ) \cdot \pi( \bm{\phi_{pred}}, \bm{\Theta}|\bm{\tilde{Z}_{1:T}} ) \partial\bm{\phi_{pred}} \partial \bm{\Theta} ,
\end{equation}
where $\bm{\tilde{Z}_{pred}}$ is predictive sample and $\bm{\phi_{pred}}$ denotes predicted process at time $t_{pred}$. We approximate (\ref{eqn:eqn16}) with posterior samples. Here, we focus on predictive posterior samples at the data scale instead of the process scale, because model performance is evaluated by comparing the predictive posterior sample and data, as described in the next section. 

\subsection{Model Evaluation}\label{sec:4.5}
Given that we are comparing predicted boundaries to true boundaries, it is reasonable to use the threat score (TS) summary measure \citep[e.g.,][]{wikle2019}
\begin{equation}
\label{eqn:eqn17}
TS=\frac{A_{11}}{A_{11}+A_{10}+A_{01}},
\end{equation}
where $A_{11}$ is the common area of overlap where an event actually occurred and is predicted to occur, and $A_{10}$ is the area where the occurrence of the event is predicted from the model but did not occur, and $A_{01}$ is the area where the event occurred but was not predicted by the model. The upper bound of $TS$ is 1 for a perfect forecast of the area of the boundary relative to the true boundary, so a TS closer to 1 indicates good model performance. In practice, when predicting signed distance functions on a grid, determining the occurrence of the event depends on a threshold parameter $\tau$. Specifically, if points on the boundary $\bm{s} \in \partial \Omega$ have a signed distance of zero $\phi(\bm{s})=0$, we set zero as our threshold parameter. That is, if $\tilde{Z}(\bm{s};t)\leq \tau=0$, we say that the event occurred at $\bm{s}$. Using posterior samples, the $TS$ summary measure can be approximated as, 
\begin{equation}
\label{eqn:eqn18}
TS^{(\ell)}=\frac{A_{11}^{(\ell)}}{A_{11}^{(\ell)}+A_{10}^{(\ell)}+A_{01}^{(\ell)}},
\end{equation}
where $TS^{(\ell)}$ is TS based on ${\ell}^{th}$ posterior sample for  $A_{11}^{(\ell)}$, $A_{10}^{\ell}$, and $A_{01}^{\ell}$. Then, we consider 
\begin{equation}
\label{eqn:eqn19}
\overline{TS}=\frac{1}{L}\sum_{\ell=1}^{L}TS^{(\ell)},
\end{equation}
as the approximation to $TS$ for the entire domain, where $L$ is the number of posterior samples drawn after burn-in.

\section{Simulation experiments}\label{sec:5}
\begin{figure}
\centering
\includegraphics[width=0.9\linewidth]{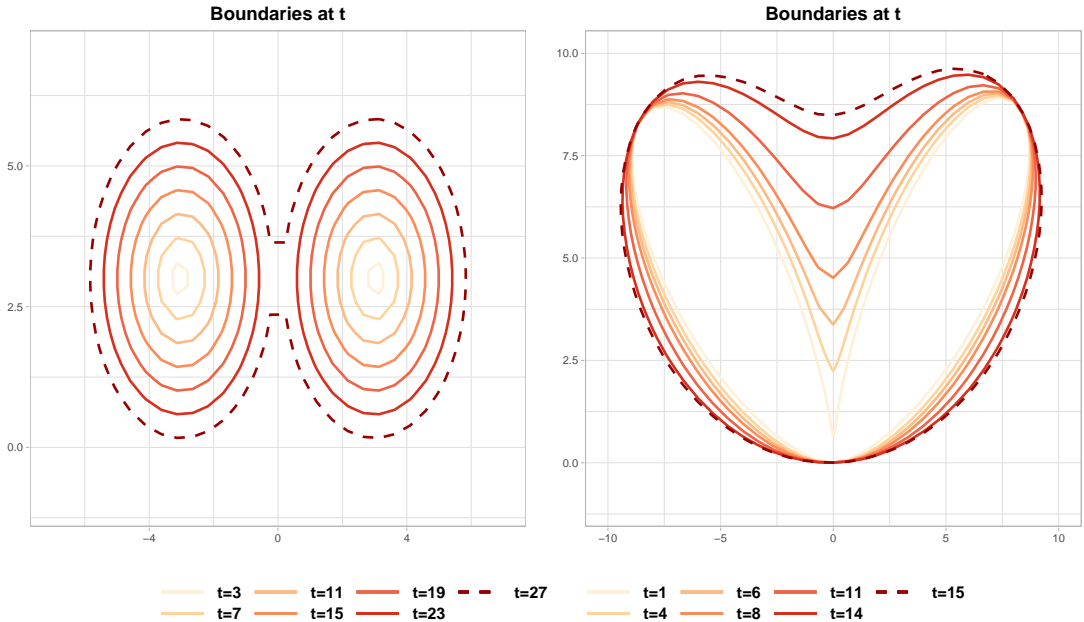}
  \caption{Numerical simulation experiments: the evolution of boundaries for each simulation. In the left figure, two boundaries evolve outward and merge into one. In the right figure, an initial "V" shaped boundary evolves to the ``north'', filling the center of the "V". In both panels, the dashed lines indicate the boundaries at $T=27$ and $15$ when a forecast is generated, respectively.}   
\label{fig3}
\end{figure}
\begin{figure}
\centering
\includegraphics[width=0.9\linewidth]{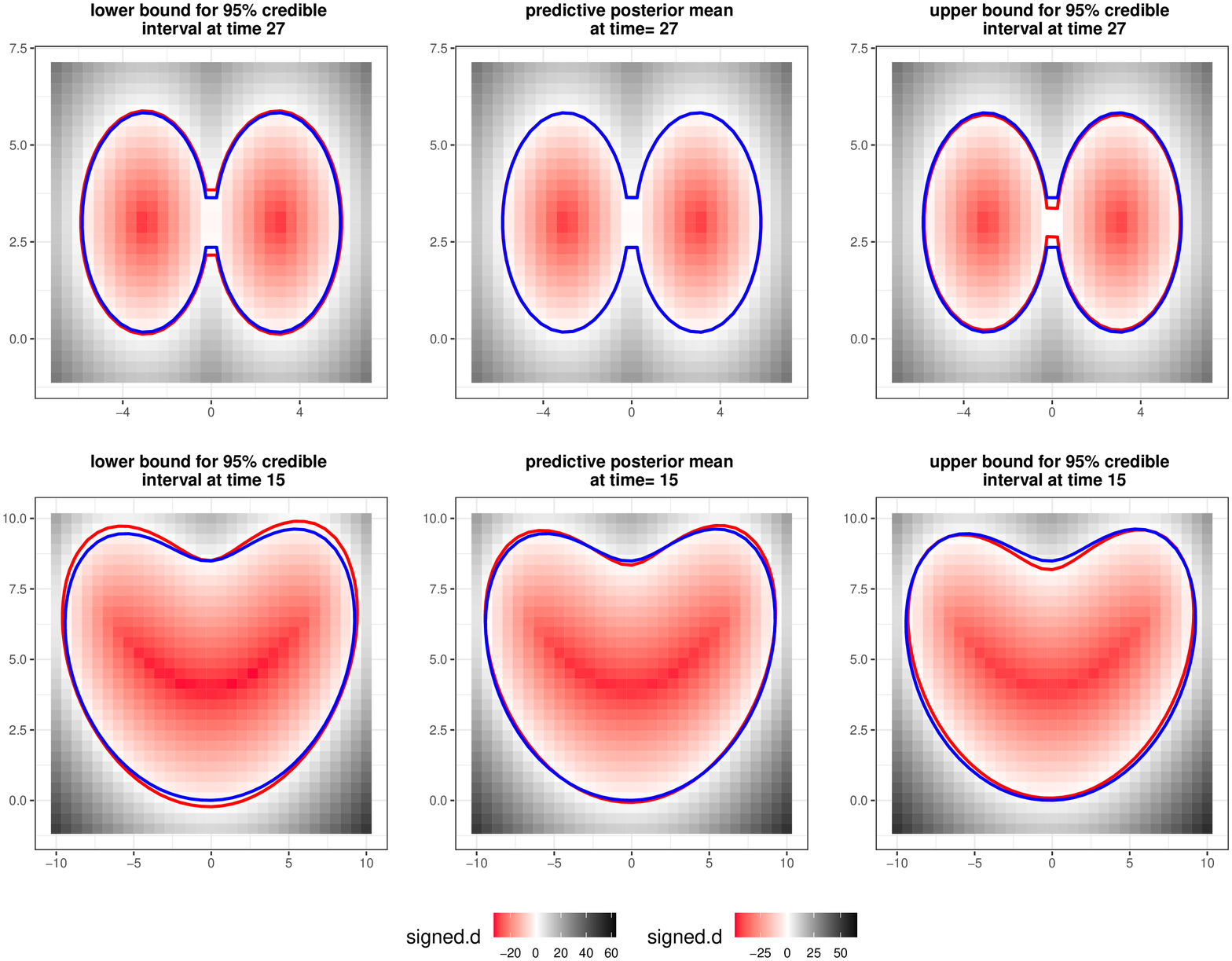}
  \caption{Numerical simulation experiments: 95\% credible intervals and posterior mean of forecasts. Figures in the first and second rows correspond to the two numerical experiments shown in Figure \ref{fig3}, respectively. The red and blue lines indicate the true and estimated boundaries, respectively, and the image map shows the corresponding signed-distance function at each pixel.}   
\label{fig4}
\end{figure}

To illustrate the methodology, we first conduct two numerical simulation experiments to demonstrate that our model is flexible enough to accommodate and forecast complicated boundary evolution. As shown in Figure \ref{fig3}, the first numerical experiment considers two distinct boundaries spreading outward and then merging into one boundary through time, while the second shows "V" shape boundaries evolving. The motivation for the first simulation is that wildfire boundaries frequently merge with other wildfire boundaries during their evolution.  The behavior of a fire discussed in \cite{hilton} inspires the second simulation. Specifically, \cite{hilton} found that initial "V" shaped fires spread in the direction of the wind, filling the center of the "V."

In the first numerical experiment, two distinct boundaries are connected at $t=27$. Obtaining a signed distance function on a $30 \times 30$ grid on the domain $[-7,7]\times[-1,7]$, we leave out data at $t=27$ and fit models with the first 26 observations using the diagonal formulation (\ref{eqn:eqn13}) with $J=6$ basis functions and without spatial varying covariates. After fitting the model with hyperparameters, $\alpha_d=\beta_d=\alpha_p=\beta_p=0.1$ and $c_{\xi}=c_{\gamma}=1000$, yielding vague priors, a one-step-ahead forecast is obtained from 30,000 posterior samples, assuming the first 20,000 as burn-in. As shown in the first row of Figure \ref{fig4}, the posterior mean for the forecast at $t=27$ shows an excellent fit to the true boundary, with a TS summary measure of 0.98. Furthermore, the 95\% credible interval plots (element-wise) satisfactorily quantify the uncertainty in these forecasts, including the true boundary. We also observed that the results using the full VAR model (\ref{eqn:eqn12}) with $J=5$ and $J=6$ basis functions (not shown) are essentially identical to those presented here.  

For the second numerical simulation experiment, the boundaries evolve to the north, filling in the center of the "V."  One can see that the initial "V" shape changes to a "heart" shape around  t=15. With a signed distance function on $31\times 31$ grid over $[-10,10]\times[-1,10]$, we investigate forecast performance at $t=15$ using the first $14$ observations to train the model. This example is more challenging than the first numerical experiment due to the fewer observations in time. With the same hyperparameters as the first experiment and using the diagonal model (\ref{eqn:eqn13}) with $J=5$ basis function, we draw 30,000 posterior samples, considering the first 20,000 as burn-in. The posterior mean and element-wise 95\% credible intervals are given in the second row of Figure \ref{fig4}.  The posterior mean of the forecast is close to the actual boundary with a TS summary measure of 0.97. As expected, given fewer data points, the 95\% credible intervals become a bit wider than in the first simulation experiment. Note also that forecasts using full VAR model (\ref{eqn:eqn12}) with $5$ basis functions give nearly identical performance with a TS summary measure 0.97.

These simulation experiments illustrate two points. First, our model can correctly forecast the progression of a fairly complex boundary with effective uncertainty quantification. In particular, the forecast from the first experiment can accommodate merging boundaries, and the second experiment can accommodate realistic fire behavior found in \cite{hilton}. Second, topological changes in boundaries can easily be represented with our method. As mentioned in Section \ref{sec:3.1}, explicit (Lagrangian) approaches to describe the boundaries cannot easily capture merging boundaries, whereas implicit methods with level sets can accommodate such structures naturally. 

\section{Wildfire applications}\label{sec:6}
We show the performance of our model on the Thomas and the Haypress fires described in Section \ref{sec:2}. A signed distance function corresponding to each fire boundary is obtained on a $30 \times 30$ grid for longitude and latitude ranges $[-119.8,-118.8] \times [34.2,34.7]$ and a $31 \times 31$ grid over $[-123.5053,-123.3252] \times [41.38407,41.54717]$ by (\ref{eqn:eqn2}), respectively. 

\subsection{The Haypress fire}\label{sec:6.1}
\begin{table}
\caption{Summary of the effect of spatial varying covariates and forecast performance for the Haypress fire at $t_{forecast,1}=2017$-$09$-$23$ 21:50:00. $\beta_{slope}$, $\beta_{aspect}$, $\beta_{canopy}$, and $\beta_{vegetation}$ correspond to coefficients for spatial varying covariates; slope, aspect, forest canopy cover, and existing vegetation cover, respectively.}
\label{table1}
\begin{adjustbox}{width=\columnwidth}
\begin{tabular}{ccccccc}
\hline
\multirow{2}{*}{Model} &   & \multicolumn{4}{c}{Posterior mean and 95\% Credible interval}                                                                                                                                                                                                                                         & \multirow{2}{*}{TS} \\ \cline{3-6}
                       & J & $\beta_{slope}$                                                        & $\beta_{aspect}$                                                          & $\beta_{canopy}$                                                         & $\beta_{vegetation}$                                                  &                     \\ \hline
$\mathcal{M}_1$        & 5 & \begin{tabular}[c]{@{}c@{}}-0.0019\\ (-0.0362, 0.0325)\end{tabular}    & \begin{tabular}[c]{@{}c@{}}0.0389\\ (0.0057, 0.0726)\end{tabular}         & \begin{tabular}[c]{@{}c@{}}-0.3690\\ (-0.4137, -0.3251)\end{tabular}     & \begin{tabular}[c]{@{}c@{}}0.1619\\ (0.1248, 0.1991)\end{tabular}     & 0.7311              \\
$\mathcal{M}_1$        & 6 & \begin{tabular}[c]{@{}c@{}}0.0257\\ (-0.0026, 0.0540)\end{tabular} & \begin{tabular}[c]{@{}c@{}}-0.0009\\ (-0.0285, 0.0269)\end{tabular} & \begin{tabular}[c]{@{}c@{}}-0.0742\\ (-0.1147, -0.0342)\end{tabular} & \begin{tabular}[c]{@{}c@{}}0.0555\\ (0.0241, 0.0865)\end{tabular} & 0.7216              \\
$\mathcal{M}_2$        & 5 & \begin{tabular}[c]{@{}c@{}}-0.0021\\ (-0.0367, 0.0321)\end{tabular}    & \begin{tabular}[c]{@{}c@{}}0.0390\\ (0.0057, 0.0726)\end{tabular}         & \begin{tabular}[c]{@{}c@{}}-0.3681\\ (-0.4125, -0.3235)\end{tabular}     & \begin{tabular}[c]{@{}c@{}}0.1609\\ (0.1237, 0.1979)\end{tabular}     & 0.7304              \\
$\mathcal{M}_2$        & 6 & \begin{tabular}[c]{@{}c@{}}0.0258\\ (-0.0020, 0.0542)\end{tabular}     & \begin{tabular}[c]{@{}c@{}}-0.0008\\ (-0.0285, 0.0264)\end{tabular}       & \begin{tabular}[c]{@{}c@{}}-0.0746\\ (-0.1149, -0.0346)\end{tabular}     & \begin{tabular}[c]{@{}c@{}}0.0554\\ (0.0240, 0.0862)\end{tabular}     & 0.7374              \\ \hline
\end{tabular}
\end{adjustbox}
\end{table}
As indicated in Section \ref{sec:1}, the primary interest here is our ability to forecast the evolution of a wildfire boundary. For this purpose, forecasting performance from models corresponding to (\ref{eqn:eqn12}) and (\ref{eqn:eqn13}) are compared for different number of basis functions, $J=5$ and $J=6$. First, we consider forecast performance at $t_{forecast,1}=2017$-$09$-$23$ 21:50:00 using the first 21 observations for training. With the hyperparameters $\alpha_d=\beta_d=\alpha_p=\beta_p=0.1$ and $c_{\beta}=c_{\xi}=c_m=c_{\gamma}=1000$, suggesting vague priors, we draw 25,000 posterior samples after discarding the first 15,000 as burn-in for one-step-ahead forecasting. Those posterior samples are then compared to the observation at $t_{forecast}=$2017$-$09$-$23 $ 21:50:00$ to evaluate model performance. Evaluation of the convergence diagnostic proposed by \citet{vats} suggests no evidence of lack of convergence.

\begin{figure}
\centering
\includegraphics[width=1\linewidth]{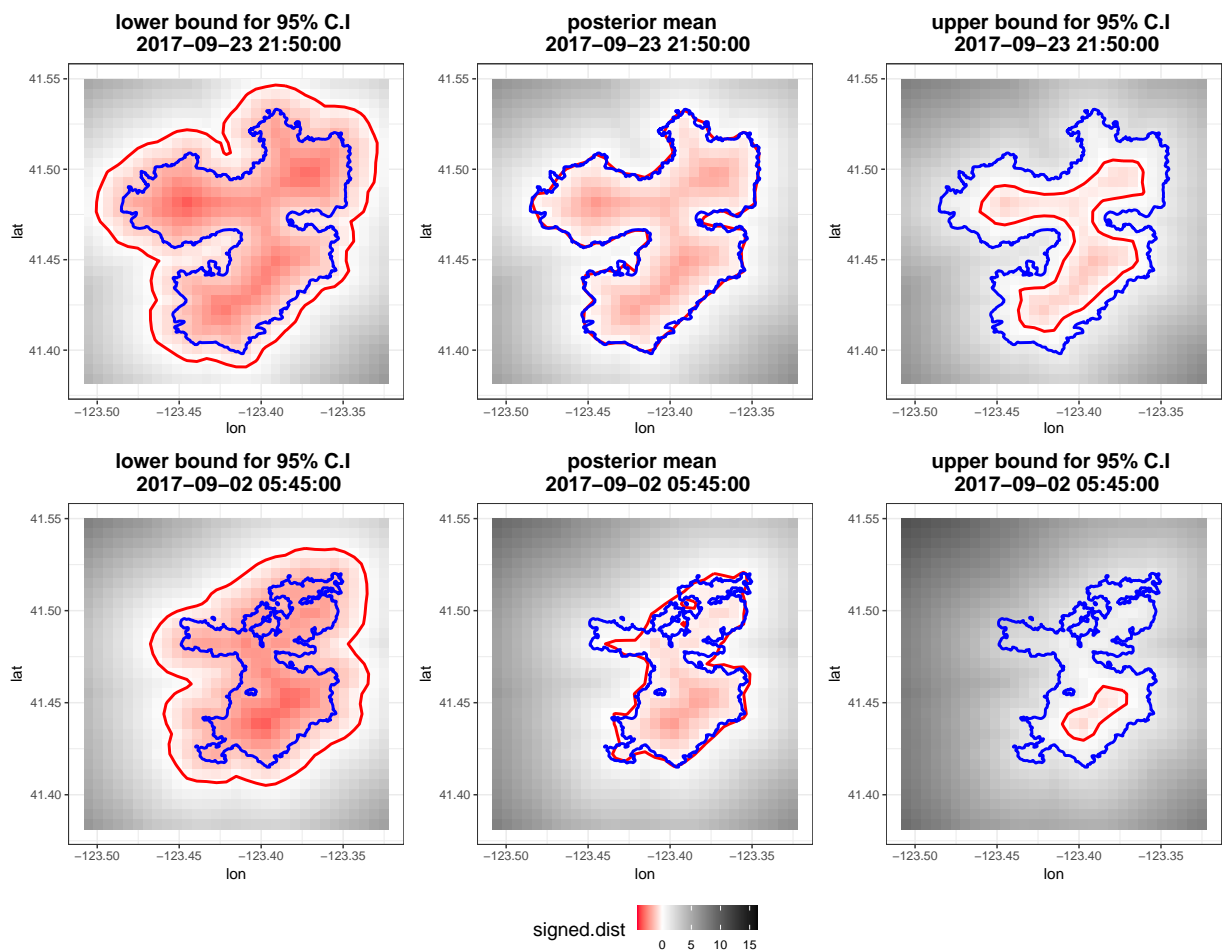}
  \caption{The Haypress fire: 95\% credible intervals and posterior mean of forecasts from best model chosen by the TS measure. Figures in first and second rows correspond to times  $t_{forecast,1}=2017\text{-}09\text{-}23$ 21:50:00 and $t_{forecast,2}=2017\text{-}09\text{-}02$ 05:45:00, respectively, with the red line indicating the forecast boundary with signed distance 0 and the blue line denoting the true boundary.}   
\label{fig5}
\end{figure}

Table~\ref{table1} summarizes the effects of spatial varying covariates and forecasting performance, where models $\mathcal{M}_1$ and $\mathcal{M}_2$ correspond to (\ref{eqn:eqn12}) and (\ref{eqn:eqn13}), respectively. Model $\mathcal{M}_2$ with 6 basis functions slightly outperforms the others in terms of the TS metric. Note that the significance of the spatially varying covariates also varies according to the number of basis functions; aspect is significant with 5 basis functions but not with 6 basis functions, while forecast canopy cover and existing vegetation cover are significant for both 5 and 6 basis functions. Looking at the result with $J=6$ basis functions, ``abundant canopy cover by life form'' could lead to more rapid wildfire propagation, whereas ``dense tree canopy cover in a stand'' deters the spread on this landscape. The posterior mean and element-wise 95\% credible intervals from the best model are depicted in the first row of Figure \ref{fig5}. The boundary from the posterior mean of the signed-distance function is very close to true boundary, and the 95\% credible intervals cover the true boundary.
\begin{table}
\caption{Summary of the effect of spatially varying covariates and forecast performance on the Haypress fire at $t_{forecast,2}=2017$-$09$-$02$ 05:45:00. $\beta_{slope}$, $\beta_{aspect}$, $\beta_{canopy}$, and $\beta_{vegetation}$ corresponds to coefficients for spatial varying covariates; slope, aspect, forest canopy cover, and existing vegetation cover, respectively.}
\label{table2}
\begin{adjustbox}{width=\columnwidth}
\begin{tabular}{ccccccc}
\hline
\multirow{2}{*}{Model} &   & \multicolumn{4}{c}{Posterior mean and 95\% Credible interval}                                                                                                                                                                                                                       & \multirow{2}{*}{TS} \\ \cline{3-6}
                       & J & $\beta_{slope}$                                                      & $\beta_{aspect}$                                                  & $\beta_{canopy}$                                                     & $\beta_{vegetation}$                                              &                     \\ \hline
$\mathcal{M}_1$        & 3 & \begin{tabular}[c]{@{}c@{}}-0.0639\\ (-0.1406, 0.0133)\end{tabular}  & \begin{tabular}[c]{@{}c@{}}0.1163\\ (0.0410, 0.1906)\end{tabular} & \begin{tabular}[c]{@{}c@{}}-0.8422\\ (-0.9358, -0.7470)\end{tabular} & \begin{tabular}[c]{@{}c@{}}0.1520\\ (0.0683, 0.2370)\end{tabular} & 0.5114              \\
$\mathcal{M}_1$        & 5 & \begin{tabular}[c]{@{}c@{}}-0.1164\\ (-0.1738, -0.0577)\end{tabular} & \begin{tabular}[c]{@{}c@{}}0.0731\\ (0.0166, 0.1295)\end{tabular} & \begin{tabular}[c]{@{}c@{}}-0.5421\\ (-0.6179, -0.4662)\end{tabular} & \begin{tabular}[c]{@{}c@{}}0.2532\\ (0.1891, 0.3174)\end{tabular} & 0.2037              \\
$\mathcal{M}_2$        & 3 & \begin{tabular}[c]{@{}c@{}}-0.0642\\ (-0.1419, 0.0135)\end{tabular}  & \begin{tabular}[c]{@{}c@{}}0.1163\\ (0.0414, 0.1920)\end{tabular} & \begin{tabular}[c]{@{}c@{}}-0.8408\\ (-0.9350, -0.7463)\end{tabular} & \begin{tabular}[c]{@{}c@{}}0.1504\\ (0.0670, 0.2331)\end{tabular} & 0.5400              \\
$\mathcal{M}_2$        & 5 & \begin{tabular}[c]{@{}c@{}}-0.1160\\ (-0.1739, -0.0581)\end{tabular} & \begin{tabular}[c]{@{}c@{}}0.0728\\ (0.0155, 0.1304)\end{tabular} & \begin{tabular}[c]{@{}c@{}}-0.5417\\ (-0.6172, -0.4655)\end{tabular} & \begin{tabular}[c]{@{}c@{}}0.2518\\ (0.1888 0.3150)\end{tabular}  & 0.5152              \\ \hline
\end{tabular}
\end{adjustbox}
\end{table}

For a second evaluation of our model, we examine forecasting performance at $t_{forecast,2}=2017$-$09$-$02$ 05:45:00 using only the first 10 observations. Given the implications of the numerical experiments, this task is much more challenging than the one investigated above for two reasons; there are much fewer time observations in this case, and there is a much more rapid rate of spread in the fire around $t_{forecast,2}$ than $t_{forecast,1}$. Table~\ref{table2} shows the performance of each model. The best model is $\mathcal{M}_2$ with 3 basis functions. The TS metric is relatively lower than in Table~\ref{table1}, but this is expected given the difficulties of this task. The lower TS value is consistent with the broader 95\% credible intervals in the second row of Figure~\ref{fig5}. In addition, note that the forecast from $\mathcal{M}_1$ with 5 basis functions is much worse than the others. We observe that this is due to the maximum modulus of $\bm{M_{\xi}}$ in (\ref{eqn:eqn12}) being greater than 1, which results in an explosive process.  That is, the model $\mathcal{M}_1$ with too many basis functions can be unstable with insufficient training data. On the other hand,  $\mathcal{M}_2$, by construction, is free of this requirement given it has a diagonal transition operator. 

\begin{figure}[t]
\centering
\includegraphics[width=0.9\linewidth]{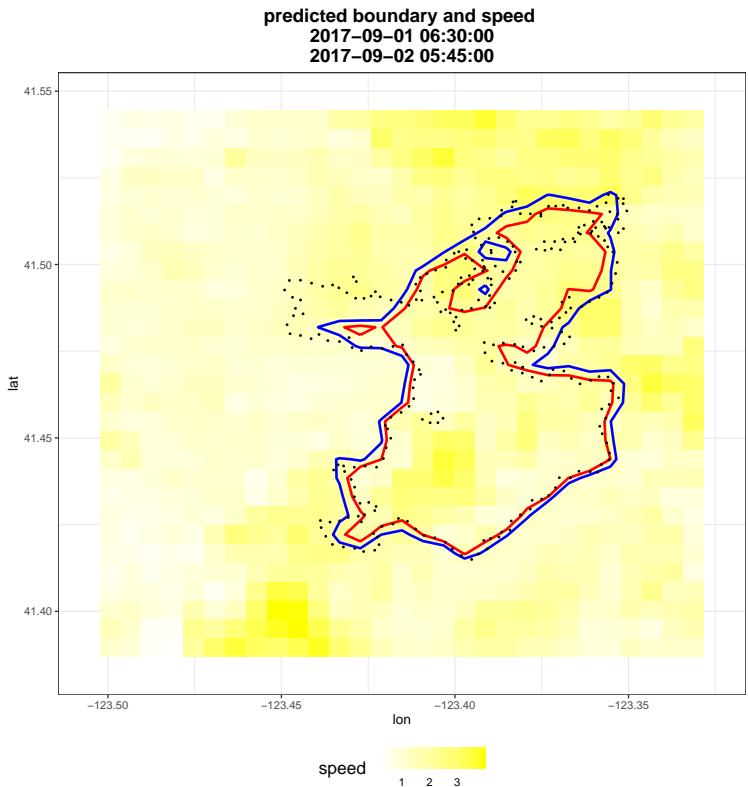}
  \caption{Posterior mean of speed and predicted boundary (red solid line) at $t=2017\text{-}09\text{-}01$ 06:30:00 and predicted boundary (blue solid line) at $t_{forecast,2}=2017\text{-}09\text{-}02$ 05:45:00. The black dotted line indicates true boundary at $t_{forecast,2}=2017\text{-}09\text{-}02$.}   
\label{fig6}
\end{figure}

Note also that the significance of the spatial varying covariates depends on the number of basis functions (perhaps due to spatial confounding), but the results are not that different between models with the same number of basis functions. Specifically, with $J=3$ basis functions, coefficients except for the slope, $\beta_{slope}$, are significant, but every coefficient is significant with $J=5$ basis functions for both models. For the best model, the ``north-facing slope'' and ``abundant canopy cover by life form'' promote wildfire propagation, whereas the higher ``dense tree canopy cover in a stand'' inhibits the spread of the wildfire for this landscape.

Figure \ref{fig6} shows the posterior mean of the speed, the boundary at $t=2017\text{-}09\text{-}01$ 06:30:00, and the boundary at $t_{forecast,2}$, respectively, from the best model. One can see that the predicted boundary at $t_{forecast,2}$ is propagated outward according to the speed at each location, as the estimated speed is greater than 0 at every location. Noticeably, a comparison between the solid red line and the dashed black line illustrates the rapid evolution in progress. In particular, the rate of spread to the west is prominent. The forecast, represented by the solid blue line, is able to capture this behavior despite having only a small number of observations in time.

\subsection{The Thomas fire}\label{sec:6.2}
\begin{figure}
\centering
\includegraphics[width=0.9\linewidth]{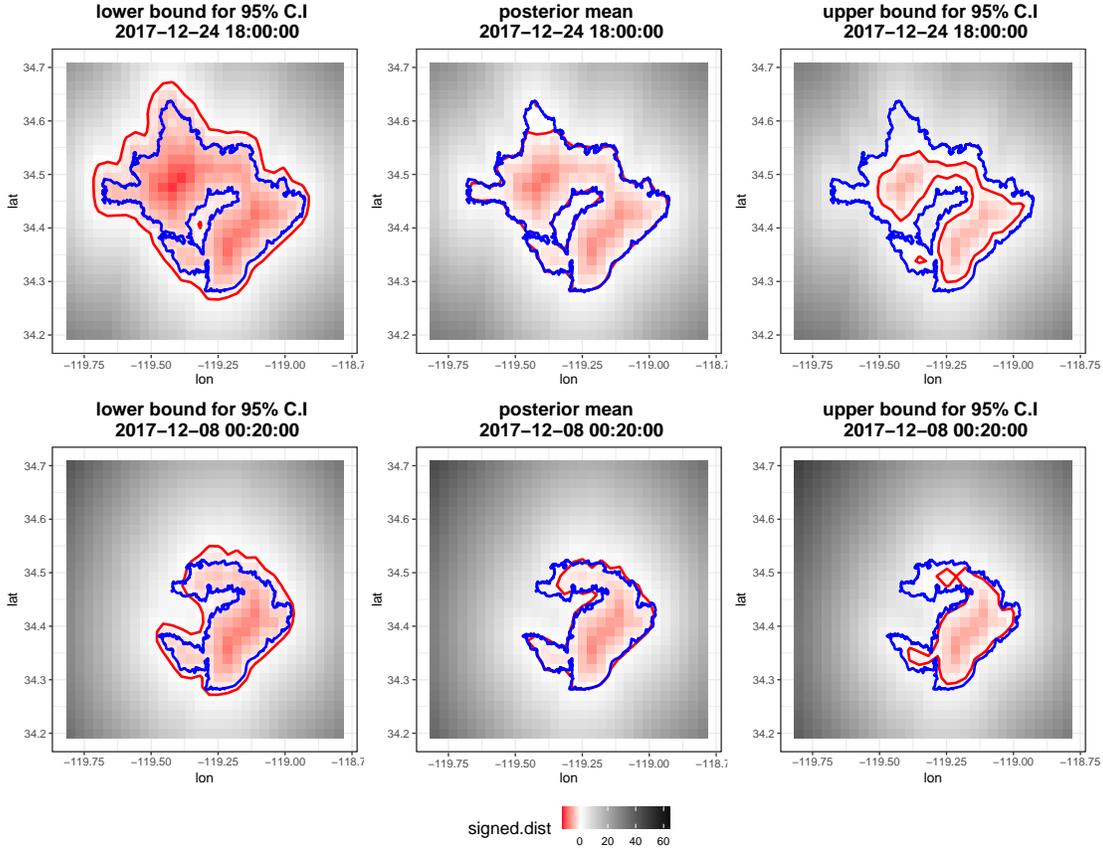}
  \caption{The Thomas fire: 95\% credible intervals and posterior mean. Figures in the first and second row correspond to a forecast at $t_{forecast,3}=2017$-$12$-$24$ 18:00:00 and a prediction at $t_{pred}=2017$-$12$-$08$ 00:20:00, respectively. The red solid line and blue solid line indicate the estimated boundary with signed distance 0 and the true boundary, respectively.}   
\label{fig7}
\end{figure}

\begin{table}
\caption{Summary of the effect of spatial varying covariates and forecast performance on the Thomas fire at $t_{forecast}=2017$-$12$-$24$ 18:00:00. $\beta_{slope}$, $\beta_{aspect}$, $\beta_{canopy}$, and $\beta_{vegetation}$ corresponds to coefficients for spatial varying covariates; slope, aspect, forest canopy cover, and existing vegetation cover, respectively.}
\label{table3}
\begin{adjustbox}{width=\columnwidth}
\begin{tabular}{ccccccc}
\hline
\multirow{2}{*}{Model} &   & \multicolumn{4}{c}{Posterior mean and 95\% Credible interval}                                                                                                                                                                                                                    & \multirow{2}{*}{TS} \\ \cline{3-6}
                       & J & $\beta_{slope}$                                                   & $\beta_{aspect}$                                                    & $\beta_{canopy}$                                                  & $\beta_{vegetation}$                                               &                     \\ \hline
$\mathcal{M}_1$        & 5 & \begin{tabular}[c]{@{}c@{}}0.2465\\ (0.0380, 0.4525)\end{tabular} & \begin{tabular}[c]{@{}c@{}}-0.0292\\ (-0.1554, 0.0942)\end{tabular} & \begin{tabular}[c]{@{}c@{}}0.3399\\ (0.2125, 0.4682)\end{tabular} & \begin{tabular}[c]{@{}c@{}}0.1930\\ (-0.0559, 0.4438)\end{tabular} & 0.7417              \\
$\mathcal{M}_1$        & 6 & \begin{tabular}[c]{@{}c@{}}0.2423\\ (0.0457, 0.4418)\end{tabular} & \begin{tabular}[c]{@{}c@{}}-0.0485\\ (-0.1637, 0.0662)\end{tabular} & \begin{tabular}[c]{@{}c@{}}0.3561\\ (0.2362, 0.4769)\end{tabular} & \begin{tabular}[c]{@{}c@{}}0.1460\\ (-0.0810, 0.3732)\end{tabular} & 0.7614              \\
$\mathcal{M}_2$        & 5 & \begin{tabular}[c]{@{}c@{}}0.2491\\ (0.0416, 0.4556)\end{tabular} & \begin{tabular}[c]{@{}c@{}}-0.0201\\ (-0.1460, 0.1055)\end{tabular} & \begin{tabular}[c]{@{}c@{}}0.3324\\ (0.2036, 0.4612)\end{tabular} & \begin{tabular}[c]{@{}c@{}}0.2239\\ (-0.0287, 0.4762)\end{tabular} & 0.7253              \\
$\mathcal{M}_2$        & 6 & \begin{tabular}[c]{@{}c@{}}0.2478\\ (0.0504, 0.4456)\end{tabular} & \begin{tabular}[c]{@{}c@{}}-0.0178\\ (-0.1347, 0.1000)\end{tabular} & \begin{tabular}[c]{@{}c@{}}0.3315\\ (0.2090, 0.4545)\end{tabular} & \begin{tabular}[c]{@{}c@{}}0.2246\\ (-0.0154, 0.4615)\end{tabular} & 0.7206              \\ \hline
\end{tabular}
\end{adjustbox}
\end{table}
We also investigate the performance of forecasting the Thomas fire at $t_{forecast_3}$= 2017$-$12$-$24 $18:00:00$ using the first 20 observations. With the same hyperparameters as used for the Haypress fire example,  25,000 posterior samples for models corresponding to ($\mathcal{M}_1$,\ref{eqn:eqn12})  and ($\mathcal{M}_2$,\ref{eqn:eqn13}) are drawn after the first 15,000 burn-in samples. There is no indication of lack of convergence based on the \citet{vats} diagnostic. Table~\ref{table3} shows the posterior summary of coefficients for the spatially varying covariates and forecast performance with $J=5,6$ basis functions. One can see that ``slope'' and ``forest canopy cover'' are significant in explaining the Thomas fire front propagation, while ``aspect'' and ``existing vegetation cover'' are not significant. This indicates that this wildfire was more likely to grow in steep terrain and in the presence of the dense forest canopy cover. We note that in this example, model $\mathcal{M}_1$ is better than model $\mathcal{M}_2$, whereas model $\mathcal{M}_2$ shows better performance than model $\mathcal{M}_1$ for the Haypress fire. This is why we consider both models in this study -- it is not always possible, {\it a priori} to determine the best model for a given fire, although it is reasonable that $\mathcal{M}_1$ performs better with limited time replicates given it has fewer parameters to estimate.  The first row of Figure \ref{fig7} shows the posterior mean and 95\% credible intervals for the best model. Given the relatively slow rate of spread and the spur in the north that develops between $t=2017$-$12$-$22$ 00:12:00 and $t=2017$-$12$-$23$ 23:10:00 in Figure \ref{fig1}, we note that the spur in the north connects somewhat suddenly at $t_{forecast,3}$. One can see that 95\% credible interval is able to capture this behavior, although the posterior mean cannot.

Even though the main focus here is on forecasting, we also investigate the performance of our model when predicting at a time $t_{pred}$, given all of the data $1 < t_{pred} < T$. Thus, we leave out the observation at $t_{pred}=2017$-$12$-$08$ 00:20:00 when fitting the model (\ref{eqn:eqn12}) and compare predictive posterior samples to observed data. From Figure \ref{fig1}, one can see that the rate of spread is quite intense around $t_{pred}$. In particular, the true boundary's upper part grows west at $t_{pred}$, then toward the north at $t$= 2017$-$12$-$08 $22:29:00$. With the same hyperparameters for forecasting and $J=6$ basis functions, 25,000 posterior samples are drawn after 15,000 burn-in samples. Again, there was no evidence of lack of convergence with the \citet{vats} diagnostic. The second row of Figure \ref{fig7} shows the 95\% credible intervals and posterior mean for this time. The predicted boundary posterior mean shows the evolution of the upper part toward the west, as in the observed data. Although the 95\% credible interval does not cover the true boundary everywhere, it has reasonable coverage and a TS value of 0.78. We also consider the performance for the model $\mathcal{M}_2$ with $J=5$ basis functions and observe that result is quite similar, with a TS value of 0.76. 
\section{Discussion and conclusion}\label{sec:7}
Representing the evolution of the boundary of interest via a signed distance function and allowing for evolution through a level set equation are valuable tools for forecasting wildfire front propagation. While this method has been implemented widely, previous applications have assumed a known, or highly parameterized rate of spread, which is a critical component in the level set method. Our approach  generalizes the rate of spread by considering it to be a stochastic process with parameters learned by previous boundary propagation and spatial covariates. By incorporating a level set equation into a hierarchical Bayesian spatio-temporal dynamic model, uncertainty can be quantified through credible intervals, and inference on environmental coefficients can be performed. Through numerical simulation experiments, we observe that our model can forecast the behavior of  complex boundaries such as one might see in large real-world wildfires. Applying our method to the Thomas and Haypress fires, we show that our model performs well in forecasting, especially when the number of data time points is not too small. Yet, with the formal uncertainty quantification, even with a small number of temporal data points, the credible intervals become wider, and the posterior mean successfully forecasts the correct pattern of wildfire front propagation. 

We formulate our model assuming evolution in the normal direction to the boundary as this is the assumption frequently made in modeling wildfires \citep{mallet, hilton,Alessandri}. We show that this assumption is reasonable with our simulated and real-world examples. However, one could formulate the model based on the more general form given in (\ref{eqn:eqn5}). Indeed, we have considered this and our experience suggests that implementation with an upwind numerical scheme requires more restrictions due to the need to satify the Courant–Friedrichs–Lewy (CFL) stability condition \citep{osher}. Additionally, it makes computation significantly more time-comsuming as we lose the conjugate full-conditionals in the MCMC sampling that is present in the model given here. 

There are multiple avenues to extend our model. First, one can consider a non-linear evolution model. Given the observed non-linear behavior in many dynamic systems \citep{chen1992}, non-linearity in the model could improve the forecasting performance \citep[e.g., see][]{hooten}. A second area of future research is concerned with the choice of the number of basis functions.  When the speed of the growth is relatively homogeneous through time, the model performance is relatively robust to the number of basis functions. However, the number of basis functions affects the model performance much during the rapid evolution with a small number of data. In particular, too many basis functions can make the model corresponding to (\ref{eqn:eqn12}) unstable. Therefore, a more principled procedure to determine the appropriate number of basis functions would be helpful. Third, it would be useful for practical applications applied to real-world fires to incorporate fire fighting effort and topological constraints into the model. The simulation model DEVS-FIRE by \cite{devs} considered fire fighting resources in their model, and it is expected that adequate fire fighting prevents the spread of wildfire. Thus, incorporating those into our model would be helpful in wildfire control management. In our applications, we used spatial varying covariates such as slope, aspect, existing vegetation cover, and forecast canopy cover as climate warming, land use, and vegetation types are known to be associated with wildfire \citep{brotons}. These partially explain the environmental impact in the rate of spread, but not entirely. For instance, topographical constraints such as large water bodies constrain the growth of wildfires. Consequently, realistic models should include more environmental and topographical features.  

Note that although we were concerned with wildfire boundary propagation here, our methodology can be applied to numerous problems that consider boundary evolution.  Examples include the growth of cancer and the retreat of Antarctic glaciers. Looking at the evolution of those boundaries, one could validate the effectiveness of treatments for cancer or elucidate the necessary attention on global warming impacts. If evolution in the normal direction is not suitable for those applications, the model based on (\ref{eqn:eqn5}) can be considered, using the upwind differencing scheme or other numerical solvers such as Hamilton-Jacobi ENO or Hamilton-Jacobi WENO scheme, introduced in \citet{osher}. One only needs to ensure that the necessary condition for the stability is satisfied in these cases.

\newpage
\appendix
\section{Appendix A: Spatial Plots for the covariates}\label{appA}
\begin{figure}[H]
\centering
\includegraphics[width=0.7\linewidth]{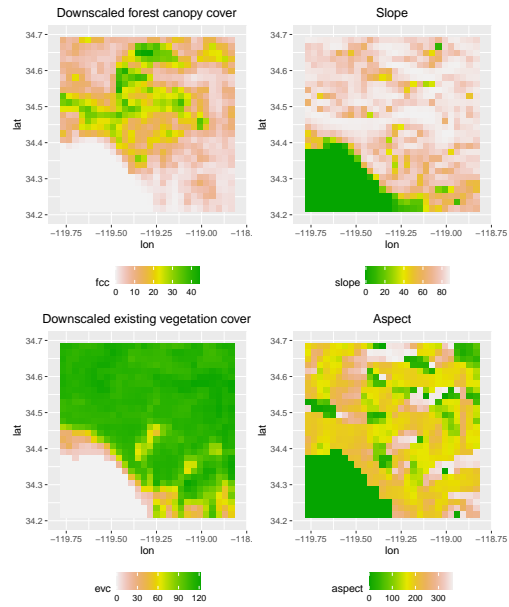}
  \caption{Spatial plots of covariates for the Haypress fire. The plots on the left show downsampled forest canopy cover and existing vegetation cover. On the right are the coresponding slope and aspect maps.}   
\label{fig8}
\end{figure}
\begin{figure}[H]
\centering
\includegraphics[width=0.7\linewidth]{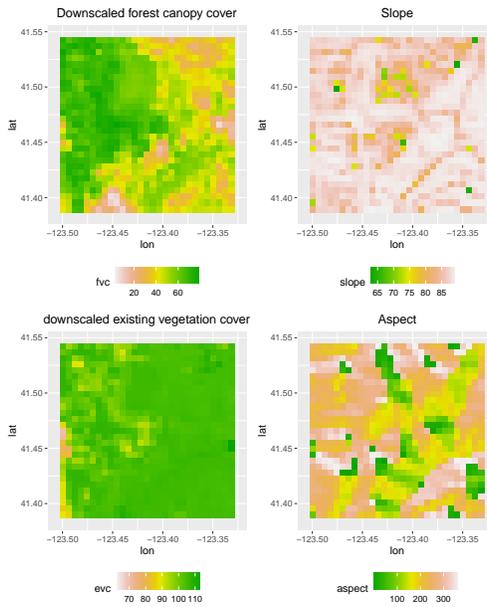}
  \caption{Spatial plots of covariates for the Thomas fire. The plots on the left show downsampled forest canopy cover and existing vegetation cover. Corresponding slope and aspect maps are shown in the right panels.}   
\label{fig9}
\end{figure}
Figure \ref{fig8} and \ref{fig9} show the spatial plots of covariates for the Thomas and Haypress fire. There are missing values on the Santa Barbara channel for the Thomas fire. Those missing values are set to zero. The analysis in this study is conducted after standardizing each covariate.

\section{Appendix B: Full conditional Distributions}\label{appB}
\subsection{Appendix B.1: Full conditional distributions for (\ref{eqn:eqn12})}\label{appB.1}
Let $T_d= \{1,1+\triangle t,\ldots,T-\triangle t,T\}$ indicate the set of time index for data, and $|T_d|$ indicate the cardinality of $T_d$. The full conditional distribution for $\sigma^2_d$ is 
\begin{align*}
\pi\bigg( \sigma^2_d|-\bigg) &\propto \prod_{t\in T_d}\pi(\bm{\tilde{Z}_t|\phi_t},\sigma^2_d)\cdot \pi \bigg(\sigma^2_d\bigg) \\
&\sim inverse.Gamma\bigg(\alpha_d+N|T_d|/2,\frac{\sum_{t\in T_d}\big(\bm{\tilde{Z}_t-\phi_t} \big)^{\top} \big(\bm{\tilde{Z}_t-\phi_t} \big)+2\beta_d}{2} \bigg),
\end{align*}
where $N$ is the number of spatial locations in the domain $\mathcal{D}\subset \mathbf{R}^2$.

Let $T_p=\{0,1, 1+\triangle t, \ldots, T-\triangle t \}$ indicate the set of time index, and $|T_p|$ indicate the cardinality of $T_p$. The full conditional distribution for $\sigma^2_p$ is 
\begin{align*}
\pi(\sigma^2_p|-)&\propto  \prod_{t\in T_p}\pi \bigg( \bm{\phi_{t+\triangle t }|\bm{\phi}_{t}},\sigma^2_p\bigg) \cdot \pi \bigg( \sigma^2_p\bigg)\\
&\sim inverse.Gamma\bigg(\alpha_p+|T_p|N/2 ,\bigg( \frac{\sum_{t\in T_p} \big(\bm{r_t})^{\top} \big( \bm{r_t} \big) +2\beta_p}{2}\bigg) \bigg),
\end{align*}
where $\bm{r_t}= \big(\bm{\phi_{t+\triangle t }}- \bm{\phi_{t}} +\bm{v_t} \triangle t \big)$.

The full conditional distribution for $\bm{\phi}_1$ is
\begin{align*}
\pi(\bm{\phi_{1}}|-)&\propto \pi(\bm{\tilde{Z}_1}| \bm{\phi_{1}})\cdot \pi(\bm{\phi_{1+\triangle t}}| \bm{\phi_{1}} ) \cdot \pi(\bm{\phi_{1}}| \bm{\phi_{0}} )\\
&\sim Gau(\bm{D}^{-1} \bm{b}, \bm{D}^{-1}),
\end{align*}
where $\bm{b}= \bigg( (\sigma^2_d \bm{I})^{-1} \tilde{\bm{Z}_1}+ (\sigma^2_p\bm{I})^{-1} \bm{\phi_{1+\triangle t}}+ (\sigma^2_p \bm{I})^{-1} \bm{v_1}\triangle t +(\sigma^2_p \bm{I})^{-1} \bm{\phi_{0}}- (\sigma^2_p\bm{I})^{-1} \bm{v_{0}} \triangle t \bigg)$ and $\bm{D}= \bigg( 1/\sigma^2_d \bm{I}+ 1/\sigma^2_p \bm{I}+1/\sigma^2_p \bm{I} \bigg)$. $\bm{\phi}_0= \tilde{Z}_1$ is assumed. 

The full conditional distribution for $\bm{\phi}_t$ when the data is missing at $t$ is 
\begin{align*}
\pi(\bm{\phi_{t}}|-)&\propto  \pi(\bm{\phi_{t+\triangle t}}| \bm{\phi_{t}} ) \cdot \pi(\bm{\phi_{t}}| \bm{\phi_{t-\triangle t}} )\\
&\sim Gau(\bm{D}^{-1}\bm{b}, \bm{D}^{-1}),
\end{align*}
where $\bm{b}= \bigg((\sigma^2_p\bm{I})^{-1} \bm{\phi_{t+\triangle t}}+ (\sigma^2_p \bm{I})^{-1} \bm{v_t}\triangle t +(\sigma^2_p \bm{I})^{-1} \bm{\phi_{t-\triangle t}}- (\sigma^2_p\bm{I})^{-1} \bm{v_{t-\triangle t}} \triangle t\bigg)$ and $\bm{D}= \bigg( 1/\sigma^2_p \bm{I}+1/\sigma^2_p \bm{I}\bigg)$.

The full conditional distribution for $\bm{\phi}_t$ when the data is not missing at $t$ is 
\begin{align*}
\pi(\bm{\phi_{t}}|-)&\propto \pi(\tilde{\bm{Z_t}}|\bm{\phi_t})\cdot  \pi(\bm{\phi_{t+\triangle t}}| \bm{\phi_{t}} ) \cdot \pi(\bm{\phi_{t}}| \bm{\phi_{t-\triangle t}} )\\
&\sim Gau(\bm{D}^{-1}\bm{b}, \bm{D}^{-1}),
\end{align*}
where $\bm{D}= \bigg( 1/\sigma^2_d \bm{I}+ 1/\sigma^2_p \bm{I}+1/\sigma^2_p \bm{I} \bigg)$ and $\bm{b}= \bigg( (\sigma^2_d\bm{I})^{-1} \tilde{\bm{Z}_t}+(\sigma^2_p\bm{I})^{-1} \bm{\phi_{t+\triangle t}}+ (\sigma^2_p \bm{I})^{-1} \bm{v_t}\triangle t +(\sigma^2_p \bm{I})^{-1} \bm{\phi_{t-\triangle t}}- (\sigma^2_p\bm{I})^{-1} \bm{v_{t-\triangle t}} \triangle t \bigg)$.

The full conditional distribution for $\bm{\phi}_T$ is 
\begin{align*}
\pi(\bm{\phi_{T}}|-)&\propto \pi(\tilde{\bm{Z_T}}|\bm{\phi_T})\cdot  \pi(\bm{\phi_{T}}| \bm{\phi_{T-\triangle t}} )\\
&\sim Gau(\bm{D}^{-1}\bm{b}, \bm{D}^{-1}),
\end{align*}
where $\bm{b}= \bigg( (\sigma^2_d\bm{I})^{-1} \tilde{\bm{Z}_T}+(\sigma^2_p \bm{I})^{-1} \bm{\phi_{T-\triangle t}}- (\sigma^2_p\bm{I})^{-1} \bm{v_{T-\triangle t}} \triangle t \bigg)$ and $\bm{D}= \bigg( 1/\sigma^2_d \bm{I}+1/\sigma^2_p \bm{I} \bigg)$.

The full conditional distribution for $\bm{\beta}$ is
\begin{align*}
\pi(\bm{\beta}|-)&\propto \prod_{t\in T_p} \pi(\bm{\phi_{t+\triangle t}}|\bm{\phi_{t}}, \bm{\beta} )\cdot \pi(\bm{\beta_a})\\
&\sim Gau(\bm{D^{-1}b},\bm{D}^{-1}),
\end{align*}
where $\bm{b}= \sum\limits_{t\in T_p}\bigg( \bm{X}^{\top} (\sigma^2_p \bm{I})^{-1}\bm{\phi_t} \triangle t -\bm{X}^{\top} (\sigma^2_p \bm{I})^{-1}\bm{\phi_{t+\triangle t}} \triangle t   -  \bm{X}^{\top} (\sigma^2_p \bm{I})^{-1} \bm{\Psi} \bm{\xi_{t}} \cdot (\triangle t)^2 \bigg) $ and $\bm{D}= \bigg( \sum\limits_{t\in T_p}(\bm{X}^{\top} (\sigma^2_p \bm{I})^{-1} \bm{X} \cdot (\triangle t)^2) +(c_{\beta}\bm{I})^{-1}   \bigg)$.

The full conditional distribution for $\bm{\xi_{0}}$ is
\begin{align*}
\pi(\bm{\xi_0}|-) &\propto \pi(\bm{\phi_{1}}|\bm{\phi_{0}}, \bm{\xi_0} )\cdot \pi(\bm{\xi_{1}}|\bm{\xi_{0}} )\cdot \pi(\bm{\xi_0})\\
&\sim Gau(\bm{D^{-1}b},\bm{D^{-1}}),
\end{align*}
where $\bm{b}= \bigg(\bm{\Psi}^{\top}  (\sigma^2_p \bm{I})^{-1}  \bm{\phi_{0}}\triangle t - \bm{\Psi}^{\top}  (\sigma^2_p \bm{I})^{-1}  \bm{\phi_{1}}\triangle t -\bm{\Psi}^{\top}  (\sigma^2_p \bm{I})^{-1} \bm{X} \bm{\beta}   (\triangle t)^2+\bm{M_{\xi}}^{\top}(\bm{\Sigma})^{-1}\bm{\xi_{1}}\bigg)$ and $\bm{D}= \bigg( \bm{\Psi}^{\top} (\sigma^2_p \bm{I})^{-1} \bm{\Psi} \cdot (\triangle t)^2  + \bm{M_{\xi}}^{\top} (\bm{\Sigma})^{-1}  \bm{M_{\xi}} + (c_{\xi}\bm{I})^{-1}\bigg)$.

The full conditional distribution for $\bm{\xi_{t}}$ for $1\leq t \leq T-2\triangle t$ is
\begin{align*}
\pi(\bm{\xi_t}|-)&\propto \pi(\bm{\phi_{t+\triangle t}}|\bm{\phi_{t}}, \bm{\xi_t} )\cdot \pi(\bm{\xi_{t+\triangle t}}|\bm{\xi_{t}} )\cdot \pi(\bm{\xi_t}|\bm{\xi_{t-\triangle t}})\\
&\sim Gau(\bm{D^{-1}b},\bm{D^{-1}}),
\end{align*}
where $\bm{D}= \bigg( \bm{\Psi}^{\top} (\sigma^2_p \bm{I})^{-1} \bm{\Psi} \cdot (\triangle t)^2  + \bm{M_{\xi}}^{\top} (\bm{\Sigma})^{-1}  \bm{M_{\xi}} + (\bm{\Sigma})^{-1}\bigg)$ and $\bm{b}= \bigg(\bm{\Psi}^{\top}  (\sigma^2_p \bm{I})^{-1}  \bm{\phi_{t}}\triangle t - \bm{\Psi}^{\top}  (\sigma^2_p \bm{I})^{-1}  \bm{\phi_{t+\triangle t}}\triangle t -\bm{\Psi}^{\top}  (\sigma^2_p \bm{I})^{-1} \bm{X} \bm{\beta}   (\triangle t)^2+\bm{M_{\xi}}^{\top}(\bm{\Sigma})^{-1}\bm{\xi_{t+\triangle t}} +(\bm{\Sigma})^{-1} \bm{M_{\xi}} \bm{\xi_{t-\triangle t}}\bigg)$.

The full conditional distribution for $\bm{\xi_{T-\triangle t}}$ is
\begin{align*}
\pi(\bm{\xi_{T-\triangle t}}|-) &\propto \pi(\bm{\phi_{T}}|\bm{\phi_{T-\triangle t}}, \bm{\xi_{T-\triangle t}} )\cdot \pi(\bm{\xi_{T-\triangle t}}|\bm{\xi_{T-2\triangle t}} )\\
&\sim Gau(\bm{D^{-1}b},\bm{D^{-1}}),
\end{align*}
where and $\bm{b}= \bigg(\bm{\Psi}^{\top}  (\sigma^2_p \bm{I})^{-1}  \bm{\phi_{T-\triangle t}}\triangle t-\bm{\Psi}^{\top}  (\sigma^2_p \bm{I})^{-1}  \bm{\phi_{T}}\triangle t - \bm{\Psi}^{\top}  (\sigma^2_p \bm{I})^{-1} \bm{X} \bm{\beta}   (\triangle t)^2 +(\bm{\Sigma})^{-1}\bm{M_{\xi}}\bm{\xi_{T-2\triangle t}} \bigg)$ and $\bm{D}= \bigg( \bm{\Psi}^{\top} (\sigma^2_p \bm{I})^{-1} \bm{\Psi}\cdot (\triangle t)^2+(\bm{\Sigma})^{-1} \bigg)$

The full conditional distribution for $\bm{m_{\xi}}=vec(\bm{M_{\xi}})$ is 
\begin{align*}
\pi(\bm{m_{\xi}}|-)\propto& \prod_{t\in T_p,t\neq T-\triangle t} \pi(\bm{\xi_{t+\triangle t}}|\bm{\xi_{t}},\bm{m_{\xi}}) \cdot \pi(\bm{m_{\xi}})\\
&\sim N(\bm{D^{-1}b},\bm{D^{-1}}),
\end{align*}
where $\bm{D}=\bigg( (\mathcal{A}_{0:T-2\triangle t}^{\top} \otimes \bm{I}_{J})^{\top} (\tilde{\bm{\Sigma}})^{-1} (\mathcal{A}_{0:T-2\triangle t}^{\top} \otimes \bm{I}_{J}) +(c_m \bm{I})^{-1} \bigg)$ and $\bm{b}= \bigg( (\mathcal{A}_{0:T-2\triangle t}^{\top} \otimes \bm{I}_{J})^{\top}  (\tilde{\bm{\Sigma}})^{-1} vec(\mathcal{A}_{1:T-\triangle t}) \bigg)$. Note that $vec(\cdot)$ is the vec operation and $\otimes$ denotes the kronecker product. Also, $\mathcal{A}_{0:T-2\triangle t}=(\bm{\xi_0},\dots,\bm{\xi_{T-2\triangle t}})$, $\bm{I_{T-\triangle t}} \otimes \bm{\Sigma_{\eta}} \equiv  \tilde{\bm{\Sigma}}$, and $\bm{I_{T-\triangle t}}$ is the $(T-1) \times (T-1)$ Identity matrix.

The full conditional distribution for $\bm{\Sigma_{\eta}}^{-1}$ is 
\begin{align*}
\pi(\bm{\Sigma^{-1}}|-) \propto &\bigg( \prod_{t\in T_p, t\neq T-\triangle t} \pi(\bm{\xi_{t+\triangle t}}|\bm{\xi_{t}},\bm{\Sigma^{-1}})\bigg) \cdot \pi(\bm{\Sigma}^{-1})\\
& \sim Wishart( \bm{D} ,|T_p|-1+ d_{\alpha}),
\end{align*}
where $\bm{D}=\bigg(  \sum\limits_{t \in T_p,t \neq T-\triangle t} \big( (\bm{\xi_{t+\triangle t}}-\bm{M_{\xi}}\bm{\xi_{t}} ) (\bm{\xi_{t+\triangle t}}-\bm{M_{\xi}}\bm{\xi_{t}} )^{\top} ) + (1000\bm{I} d_{\alpha}\big) \bigg)^{-1}$ and $d_{\alpha}=J-1$.
\subsection{Appendix B.2: Full conditional distributions for (\ref{eqn:eqn13})}\label{appB.2}
The full conditional distribution for $\bm{\gamma}$ is 
\begin{align*}
\pi(\bm{\gamma}|-)\propto& \prod_{t\in T_p,t \neq T-\triangle t} \pi(\bm{\xi_{t+\triangle t}} |\bm{\xi_{t}}, \bm{\gamma}) \cdot \pi(\bm{\gamma})\\
&\sim N(\bm{D^{-1}b},\bm{D^{-1}}),
\end{align*}
where $\bm{b}= \bigg( \sum\limits_{ \substack{t \in T_p \\ t\neq T-\triangle t}} diag(\bm{\xi_{t}})^{\top}\bm{\Sigma^{-1}} \bm{\xi_{t+\triangle t}}\bigg)$ and $\bm{D}= \bigg( \sum\limits_{ \substack{t \in T_p \\ t\neq T-\triangle t}} \big( diag(\bm{\xi_{t}})^{\top}\bm{\Sigma^{-1}} diag(\bm{\xi_{t}})\big) + (c_{\gamma}\bm{I})^{-1} \bigg)$. 

The full conditional distributions for $\bm{\xi_t}$ and $\bm{\Sigma^{-1}}$ can be obtained by replacing $\bm{M_\xi}$ with $diag(\bm{\gamma})$ in the Appendix \ref{appB.1}.

\bibliographystyle{apalike}
\bibliography{references}

\end{document}